\documentclass[cameraready]{Interspeech}

\title{AI-based Cognitive-linguistic Features for Dementia Assessment in Picture Description}

\author[affiliation={1}, orcid=0000-0002-6554-9438, correspondingauthor]{Lingfeng}{Xu}
\author[affiliation={1,2}]{Prad}{Kadambi}
\author[affiliation={1}]{Samuel}{Goldinger}
\author[affiliation={1,2}]{Visar}{Berisha}
\author[affiliation={3}]{Kimberly D.}{Mueller}
\author[affiliation={1}]{Julie}{Liss}

\address{
    $^1$ College of Health Solutions, Arizona State University, USA \\
    $^2$ School of Electrical, Computer and Energy Engineering, Arizona State University, USA \\
    $^3$ Department of Communication Sciences and Disorders, University of Wisconsin-Madison, USA 
}

\email{\{lingfen3, pkadambi, sagoldin, visar, jmliss\}@asu.edu, kdmueller@wisc.edu}


\usepackage{comment}

\begin{document}

\maketitle

\begin{abstract}
    Picture descriptions provide valuable insights into several clinical constructs related to cognitive-linguistic abilities. However, operationalizing these constructs into quantitative measures remains challenging, limiting interpretability and clinical utility. We introduced seven constructs tailored to the Cookie Theft picture description task and prompted large language models (LLMs) to evaluate them, generating severity scores and example-based explanations. Among the examined LLMs, Claude 3.5 Sonnet performed the best, producing severity scores that significantly distinguish cognitively impaired individuals from healthy controls. The model achieves a high accuracy of 85\% on the ADReSS dataset. Expert evaluation of Claude’s scores and explanations yields a 3.99/5 average agreement. The findings demonstrate the potential of LLMs to operationalize clinical constructs and generate interpretable evaluations, offering a promising approach for accessible cognitive screening tools.

\end{abstract}

\section{Introduction}

Cognitive impairment is a critical early indicator of several neurodegenerative diseases \cite{morris2001mild, hoops2009validity, duff2010mild}. Effective identification enables timely diagnosis, accurate risk assessment, and early interventions that can slow disease progression and improve patient outcomes \cite{dubois2015timely, van2023lecanemab}. Neuropsychological assessments, many of which include speech-language tasks, are used to diagnose cognitive impairment. The picture description task, in which one describes in detail what is happening in a picture, is a common component used to assess expressive language skills \cite{cummings2019describing}. 

Analysis of narratives elicited through picture description provides insights into cognitive function across several clinical constructs \cite{cummings2019describing, mueller2018connected}, including lexical richness, grammatical complexity, discourse organization, and coherence. In digital health applications, researchers have developed methods to automatically extract acoustic \cite{PULIDO2020113213, themistocleous2020voice, wang2022speech, xu2025articulatory} and linguistic \cite{verfaillie2019high, mueller2018declines} features from picture descriptions to distinguish cognitively unimpaired individuals from those with impairments. High-dimensional feature representations generated by artificial intelligence (AI) models \cite{zhu21e_interspeech, roshanzamir2021transformer, yuan20_interspeech} have also demonstrated classification capabilities. While the reported accuracy of these models seems promising, existing studies have several limitations. First, many clinical constructs are inherently complex and not easily operationalized \cite{liss2024operationalizing}, making it difficult to measure and generate normative data for meaningful comparison across participant groups. Second, the task generates a large amount of data, challenging manual transcription and analysis \cite{cummings2019describing}. Third, traditional speech and language features may relate to the clinical constructs, but they were not designed to measure these constructs, making them harder to interpret in a clinical setting. Finally, many effective features \cite{zhu21e_interspeech, yancheva2015using, li2021comparative} are high-dimensional, which not only exacerbate interpretability challenges, but also raise concerns regarding their repeatability \cite{stegmann2020repeatability}.

Recently, the advancement of large language models (LLMs), including ChatGPT \cite{achiam2023gpt}, Claude \cite{anthropic_claude35_modelcard_2024}, and LLaMA \cite{touvron2023llama}, offers new promise for digital health applications. These models perform well in few-shot scenarios \cite{nori2023capabilities} and provide human-readable rationales \cite{wang2023can}, making them well-suited for clinical settings. Additionally, recent studies have shown that LLMs can operationalize abstract clinical constructs (e.g. verbal fluency, syntactic complexity, semantic density, discourse coherence) into quantitative measures for picture description-based cognitive status evaluation, bridging clinical theory and computational assessment \cite{wang2023text, bang2024alzheimer, botelho24_interspeech, chen2024profiling}. A typically adopted pipeline is providing construct definitions to the LLM and asking the model to assign numerical scores to each construct. These scores are either directly analyzed for their discriminative ability \cite{wang2023text} or used as features in downstream classification models \cite{bang2024alzheimer, botelho24_interspeech, chen2024profiling}. Despite initial promise, several problems remain. First, existing works often relied on general-purpose constructs not tailored to picture description tasks. Considering that the task context can substantially alter language use \cite{nyongesa2025artificial}, such generic features may be insufficient. A task-specific evaluation grounded in picture description content, such as object prioritization, temporal sequencing, and mental state attribution, is optimized for identifying cognitive impairment. Second, instead of prompting LLMs to directly operationalize clinical constructs, many studies still define them using basic numerical features like word or phrase frequency \cite{wang2023text} and sentence length \cite{botelho24_interspeech}, which may not align well with the constructs they intend to represent. Moreover, LLMs often struggle with counting \cite{zhang2024counting} and arithmetic \cite{mcleish2024transformers} tasks without specialized guidance. Finally, current research largely focuses on proprietary models like ChatGPT, with limited exploration of open-source ones.

This study aims to address these research gaps by developing an LLM-based cognitive status evaluation framework specifically tailored to picture description tasks. We focus on the widely used Cookie Theft picture description task \cite{goodglass2001bdae} and draw on the work of Cummings \cite{cummings2019describing}, who systematically analyzed this task and identified a comprehensive set of task-specific constructs to assess sources of cognitive-linguistic breakdown. These include: 1) Saliency of information; 2) Semantic categories; 3) Referential cohesion; 4) Causal and temporal relations; 5) Mental state language; 6) Structural language and speech; and 7) General cognition and perception. These constructs are inherently complex and are difficult to operationalize using traditional natural language processing methods, highlighting the need for LLMs. Our experiments demonstrate that, with a small number of training samples, LLMs can directly evaluate each construct and assign a discriminative score that reflects the severity of deficit. Additionally, the models generate natural language explanations to support their assessments, including direct quotations from the input transcript that highlight the specific segments where deficits are identified. In short, this study leverages LLMs to enable clinically meaningful evaluation of cognitive impairment, contributing to the development of scalable, interpretable, and accessible screening tools for real-world clinical applications.

\section{Methods}

\subsection{Data preparation}

This study used data from the Pitt corpus of DementiaBank \cite{becker1994natural} and the Wisconsin Alzheimer’s Disease Research Center (W-ADRC) clinical core \cite{asthana2024wisconsin}. All participants completed the standard Cookie Theft picture description task from the Boston Diagnostic Aphasia Examination \cite{goodglass2001bdae}. Manual transcripts were preprocessed into clean paragraphs for LLM input. During this step, we removed most annotations originally added by clinical experts. These annotations primarily highlight linguistic phenomena of interest for clinical analysis (e.g. grammatical errors, semantic information units, and repetitions), which should be identified by LLMs from the texts alone. The only annotations retained were those indicating unfilled pauses, as such temporal features cannot be inferred solely from the textual transcripts. Next, we selected samples with available Mini-Mental State Examination (MMSE) scores \cite{folstein2010mini}, a widely used measure of global cognitive function ranging from 0 to 30, with higher scores indicating better cognitive performance. The selected samples were categorized into Control and Clinical groups based on their diagnostic labels. For the DementiaBank dataset, samples labeled ``Control" were assigned to the Control group, with all others assigned to the Clinical group. In the W-ADRC dataset, samples labeled ``Normal" were placed in the Control group, and the rest in the Clinical group. To prevent speaker-level data leakage, participants who transitioned from a control to a clinical cognitive status were included only in the Clinical group, with only their clinical samples retained.

In addition to manual transcripts, we used WhisperX \cite{bain23_interspeech} to automatically transcribe the speech recordings. The resulting automatic speech recognition (ASR)–generated transcripts served as another input source for the LLMs. As an enhanced version of Whisper, WhisperX provides faster transcription with accurate speaker diarization and word-level timestamps. All automatic transcriptions in this study were generated using the WhisperX-large-v2 model. Raw audio recordings were processed through the model, and the word-level timestamps were used to identify unfilled pauses. To maintain consistency with manual annotations, pauses between 1–3 seconds were labeled as ``$<$short pause$>$", those between 3–5 seconds as ``$<$medium pause$>$", and those exceeding 5 seconds as ``$<$long pause$>$" in the ASR-generated transcripts. Table \ref{tab:dataset} presents the demographics, sample sizes, and MMSE scores for each dataset. Throughout the paper, the Pitt corpus from DementiaBank is referred to simply as DementiaBank for brevity. For W-ADRC, Rey Auditory Verbal Learning Test \cite{rey1941examen} total scores (RAVLT total), which measure learning and memory abilities, are also provided.

\begin{table*}[th]
\caption{An overview of datasets used in this study}
\label{tab:dataset}
\centering
\begin{tabular}{l c c c c}
\toprule
\textbf{Dataset}            & \multicolumn{2}{c}{\textbf{DementiaBank}} & \multicolumn{2}{c}{\textbf{W-ADRC}}      \\
\midrule
Label              & Clinical        & Control        & Clinical       & Control        \\
Number of subjects & 191             & 96             & 49             & 184            \\
Number of samples  & 277             & 182            & 52             & 200            \\
Age (year)         & 71.04 ±   8.63  & 63.60 ±   8.03 & 74.34 ±   9.69 & 67.83 ±   8.52 \\
Education (year)   & 12.25 ±   2.91  & 13.91 ±   2.55 & 15.22 ±   2.87 & 16.44 ±   2.21 \\
Number of females  & 124             & 57             & 20             & 123            \\
MMSE               & 19.71 ±   5.65  & 29.13 ±   1.11 & 25.81 ±   3.25 & 29.52 ±   0.92 \\
RAVLT total        & -               & -              & 28.93 ± 14.45  & 52.36 ± 10.39  \\
\bottomrule
\end{tabular}
\end{table*}

\subsection{Task-specific constructs for cognitive status evaluation}

Seven clinical constructs tailored to the Cookie Theft picture description task were introduced to support comprehensive evaluation of cognitive-linguistic abilities \cite{cummings2019describing}. Brief definitions are provided in Table \ref{tab:construct}. A structured prompt was developed to guide the LLMs in evaluating each clinical construct, assigning severity scores using a 4-point scale, and generating example-based explanations. 

\begin{table*}[th]
\caption{A summary of the proposed task-specific constructs}
\label{tab:construct}
\centering
\begin{tabular}{l l}
\toprule
\textbf{Construct}               & \textbf{Definition}                                                                                                                                   \\
\midrule
Saliency of information          & The ability to prioritize and describe the most relevant elements (e.g. boy, girl) of the picture.                                   \\
Semantic categories              & The specificity of vocabulary (e.g. boy vs. child) used in descriptions.                                                                            \\
Referential cohesion             & The clarity and consistency of pronoun usage (e.g. he, she) to refer to characters and objects                             \\
& in the picture.    \\
Causal and temporal relations    & The ability to describe events in a logical sequence and temporal order (e.g. the sink is\\
& overflowing because the mother left the tap running).\\
Mental state language            & The extent to which a speaker infers the thoughts, intentions, and emotions of the characters          \\
& (e.g. the mother is not paying attention).\\
Structural language and speech   & The accuracy and complexity of a speaker’s phonology, syntax, and semantics.                                                                        \\
General cognition and perception & The speaker’s overall cognitive functioning, shown in how the picture is perceived, organized,     \\
& and described without repetition.\\
\bottomrule
\end{tabular}
\end{table*}

\begin{figure}[t]
  \centering
  \includegraphics[width=\linewidth]{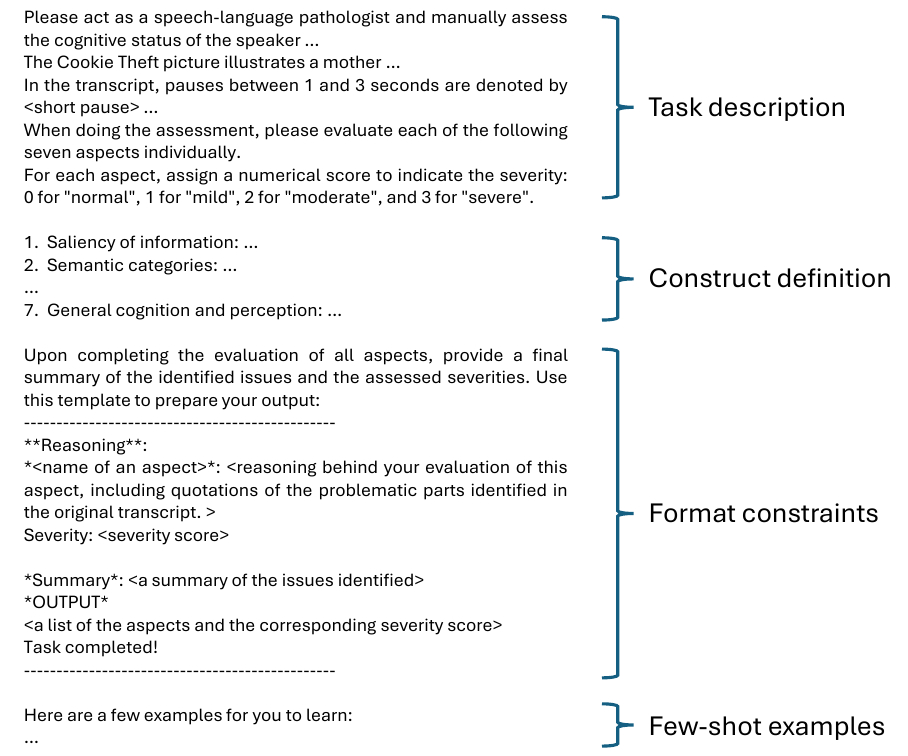}
  \caption{The prompt used for LLM-based cognitive evaluation}
  \label{fig:prompt}
\end{figure}

As illustrated in Figure \ref{fig:prompt}, the prompt begins with a detailed task description. The first paragraph defines the role of the LLM and its intended audience, and outlines the primary objective: to assess speakers’ cognitive status by analyzing their descriptions of the Cookie Theft picture. The second paragraph provides background information on the Cookie Theft image, followed by a clarification of the pause annotation format used in the transcripts. The final paragraph specifies the core task requirements. The LLMs are instructed to evaluate each input transcript across seven clinical constructs and assign a severity score to each of them, using a scale of 0 (“normal”), 1 (“mild”), 2 (“moderate”), and 3 (“severe”).

Following the task description, the definitions of the clinical constructs are provided to the LLMs. Further constraints are then introduced under the “Format constraints” section in Figure \ref{fig:prompt}. In addition to assigning severity scores, the LLMs are required to generate a paragraph summarizing the identified cognitive issues. Both the severity scores and the summary must be organized according to the template specified in the prompt. For each clinical construct, the LLMs must provide a rationale for the assigned score, supported by direct quotations from the input transcript. The use of a standardized template is intended to provide more precise guidance to the models.

Finally, five examples are included in the prompt to support few-shot learning by the LLMs. These examples were derived from Cummings' original study \cite{cummings2019describing}, in which the author conducted an in-depth analysis of five DementiaBank samples to illustrate the applicability of the seven proposed clinical constructs. The detailed evaluations presented in that work were extracted and treated as reference standards for our examples. Severity scores were assigned based on the author’s assessments and reviewed by a senior speech-language pathologist on our team to ensure clinical accuracy. As these examples were used for few-shot learning, they were excluded from all subsequent evaluation experiments in this study to avoid data leakage \cite{kapoor2022leakage}.

\subsection{Experimental settings}

The experiments were conducted on GPT-4o-mini-2024-07-18, GPT-4o-2024-08-06, Claude 3.5 Sonnet-20241022, and LLaMA-3.2-3B-instruct. The selected proprietary models were the highest-performing options within our budget constraints at the time of experimentation. The inclusion of an open-source small-scale model enabled the evaluation of a locally hosted system in clinical settings and allowed the experiments to be extended to private data (i.e., W-ADRC data). The temperature was set to 0.2 to introduce minimal variability while preserving output consistency. 

While the severity scores generated by each out-of-the-box LLMs were evaluated, further adaptation strategies were explored. For one thing, we fine-tuned the LLaMA-based model using QLoRA \cite{dettmers2023qlora} by training a 4-bit adapter. The adaptation was performed on the DementiaBank dataset, with output severity scores provided by Claude, our best-performing LLM. Teacher forcing was applied during training. Since the LLM outputs contained substantially more non-numerical tokens (corresponding to explanatory text) than numerical tokens (corresponding to severity scores), the loss associated with numerical tokens was upweighted by a factor of 10. Training was conducted on a desktop equipped with an Intel i9-9900K CPU and an NVIDIA RTX 3090 GPU with 24 GB of VRAM. The model was trained for three epochs using a learning rate of 2e-4 and a batch size of 16. Evaluation was performed on the W-ADRC dataset using a temperature of 0.2 and a maximum sequence length of 2048.

Another adaptation strategy was training a logistic regression model on DementiaBank to predict the severity scores. The severity scores generated by Claude were used as the ground-truth labels for model training. Manual or ASR-generated transcripts were first converted into high-dimensional embeddings using a pre-trained Bidirectional Encoder Representations from Transformers (BERT) model \cite{devlin2019bert}, and then used as inputs to the logistic regression model to estimate each severity score. The trained model was also evaluated on the W-ADRC dataset, enabling a comparison across adaptation strategies. 
Unlike the teacher forcing approach, which retains next-token prediction as the target task, logistic regression allows explicit specification of the target task (severity score estimation) and is therefore expected to yield more accurate results.

Once outputs were obtained from all models and strategies, the effects of age, gender, and education were corrected. Welch’s $t$-test and Hedge's $g$ were then used to evaluate the ability of the scores to differentiate between Clinical and Control groups, with larger Hedge's $g$ values indicating stronger effect sizes. Bonferroni correction was used for multiple test corrections. Pearson correlation coefficient (PCC) was computed to evaluate alignment with existing cognitive scores (e.g. MMSE and RAVLT total). For test–retest reliability, participants with two visits and the same diagnosis were analyzed using two test-retest reliability metrics \cite{stegmann2020repeatability}, Intraclass Correlation Coefficient (ICC) and Within-Subjects Coefficient of Variation (WSCV). Higher ICC and lower WSCV indicate better reliability. To aid interpretation of WSCV, the percentage difference between groups was also reported, which was calculated as the absolute mean difference divided by the Control group’s mean.

\begin{table*}[th]
\caption{Performance of the LLMs on DementiaBank using manual transcripts}
\label{tab:DB_manual}
\centering
\begin{tabular}{@{}l l l l l l l l l}
\toprule
\textbf{Construct} & \textbf{LLM} & \textbf{Control} & \textbf{Clinical} & \textbf{Group} & \textbf{Hedge’s $g$} & \textbf{PCC} & \textbf{ICC} & \textbf{WSCV} \\
& & \textbf{(mean)} & \textbf{(mean)} & \textbf{Difference} & & & & \\
\midrule
\textbf{Salience of information}          & Claude       & 0.60                    & 1.57                     & 1.61                      & $-1.15^{**}$            & $-0.65^{**}$      & $0.55^{**}$       & 0.45          \\
                                          & 4o           & 0.96                    & 1.72                     & 0.79                      & $-0.96^{**}$            & $-0.60^{**}$      & $0.48^{**}$       & 0.36          \\
                                          & 4o-mini      & 1.87                    & 2.08                     & 0.11                      & $-0.50^{**}$            & $-0.38^{**}$      & 0.19         & 0.10          \\
                                          & LLaMA        & 1.78                    & 2.11                     & 0.19                      & $-0.35^{**}$            & $-0.19^{**}$      & 0.12         & 0.41          \\
\textbf{Semantic   categories}            & Claude       & 0.65                    & 1.51                     & 1.32                      & $-1.05^{**}$            & $-0.62^{**}$      & $0.53^{**}$       & 0.49          \\
                                          & 4o           & 1.22                    & 1.83                     & 0.49                      & $-0.88^{**}$            & $-0.56^{**}$      & $0.43^{**}$       & 0.29          \\
                                          & 4o-mini      & 1.59                    & 1.94                     & 0.22                      & $-0.60^{**}$            & $-0.46^{**}$      & $0.36^{**}$       & 0.15          \\
                                          & LLaMA        & 1.47                    & 1.87                     & 0.27                      & $-0.38^{**}$            & $-0.25^{**}$      & 0.05         & 0.61          \\
\textbf{Referential cohesion}             & Claude       & 0.37                    & 1.27                     & 2.43                      & $-0.88^{**}$            & $-0.62^{**}$      & $0.41^{**}$       & 0.52          \\
                                          & 4o           & 0.51                    & 1.42                     & 1.78                      & $-0.95^{**}$            & $-0.60^{**}$      & $0.53^{**}$       & 0.43          \\
                                          & 4o-mini      & 1.69                    & 1.98                     & 0.17                      & $-0.52^{**}$            & $-0.44^{**}$      & $0.31^{**}$       & 0.15          \\
                                          & LLaMA        & 2.09                    & 2.44                     & 0.17                      & $-0.35^{**}$            & $-0.22^{**}$      & 0.15         & 0.37          \\
\textbf{Causal   and temporal relations}  & Claude       & 1.06                    & 1.75                     & 0.66                      & $-0.71^{**}$            & $-0.50^{**}$      & $0.43^{**}$       & 0.50          \\
                                          & 4o           & 1.79                    & 2.25                     & 0.26                      & $-0.57^{**}$            & $-0.41^{**}$      & $0.42^{**}$       & 0.25          \\
                                          & 4o-mini      & 2.02                    & 2.12                     & 0.05                      & $-0.30^{**}$            & $-0.27^{**}$      & 0.02         & 0.08          \\
                                          & LLaMA        & 1.68                    & 2.04                     & 0.21                      & $-0.29^{**}$            & $-0.22^{**}$      & 0.02         & 0.49          \\
\textbf{Mental state language}            & Claude       & 1.48                    & 2.05                     & 0.38                      & $-0.53^{**}$            & $-0.39^{**}$      & $0.49^{**}$       & 0.37          \\
                                          & 4o           & 2.24                    & 2.59                     & 0.15                      & $-0.29^{**}$            & $-0.29^{**}$      & $0.39^{**}$       & 0.29          \\
                                          & 4o-mini      & 2.29                    & 2.50                     & 0.09                      & $-0.31^{**}$            & $-0.25^{**}$      & $0.23^{*}$       & 0.19          \\
                                          & LLaMA        & 1.60                    & 1.91                     & 0.19                      & $-0.17$            & $-0.21^{**}$      & 0.07         & 0.59          \\
\textbf{Structural language and speech}   & Claude       & 1.11                    & 1.96                     & 0.76                      & $-0.95^{**}$            & $-0.62^{**}$      & $0.50^{**}$       & 0.34          \\
                                          & 4o           & 1.08                    & 1.80                     & 0.67                      & $-0.95^{**}$            & $-0.60^{**}$      & $0.46^{**}$       & 0.31          \\
                                          & 4o-mini      & 1.71                    & 2.01                     & 0.18                      & $-0.53^{**}$            & $-0.44^{**}$      & $0.36^{**}$       & 0.13          \\
                                          & LLaMA        & 1.84                    & 2.23                     & 0.21                      & $-0.34^{**}$            & $-0.25^{**}$      & 0.19         & 0.44          \\
\textbf{General cognition and perception} & Claude       & 0.94                    & 1.96                     & 1.09                      & $-1.04^{**}$            & $-0.61^{**}$      & $0.53^{**}$       & 0.45          \\
                                          & 4o           & 1.39                    & 2.08                     & 0.49                      & $-0.65^{**}$            & $-0.53^{**}$      & $0.52^{**}$       & 0.36          \\
                                          & 4o-mini      & 1.93                    & 2.10                     & 0.09                      & $-0.36^{**}$            & $-0.37^{**}$      & 0.10         & 0.10          \\
                                          & LLaMA        & 1.81                    & 2.17                     & 0.20                      & $-0.30^{**}$            & $-0.20^{**}$      & 0.06         & 0.55          \\
\bottomrule
\multicolumn{9}{l}{\footnotesize $^{**}p < 0.01$, $^{*}p < 0.05$} \\
\end{tabular}
\end{table*}

\subsection{A machine learning-based dementia detection task}

To benchmark the generated scores, an eXtreme Gradient Boosting (XGBoost) model \cite{10.1145/2939672.2939785} was trained to classify dementia patients and healthy controls in the ADReSS \cite{luz20_interspeech} dataset. ADReSS is a balanced subset of the DementiaBank dataset \cite{becker1994natural}, with 156 participants (78 with Alzheimer’s disease, 78 controls). Following the original split, 108 participants were used for training and 48 for testing. During data preparation, we identified that all but one sample in the ADReSS dataset overlapped with those used in our study. The single non-overlapping sample was processed using the same pipeline to ensure consistency. Additionally, two training samples were found to be part of the five examples used for few-shot learning. These samples were excluded from the experiment to avoid data leakage.

In line with existing studies \cite{bang2024alzheimer, botelho24_interspeech}, we adopted a nested cross-validation strategy with majority voting for model development. Specifically, a ten-fold cross-validation was performed on the training set. Within each fold, an inner five-fold cross-validation was conducted to fine-tune the XGBoost model. The ten resulting fine-tuned models were then evaluated on the same held-out test set, and their predictions were aggregated using majority voting to generate the final classification outcomes. The hyperparameters fine-tuned included the learning rate, number of trees, maximum tree depth, minimum child weight, and minimum loss reduction required to make a split. 

The classification performance was evaluated using accuracy, precision, recall, and F1 score, a widely used metric that balances precision and recall on a 0–1 scale, with higher values indicating superior performance. SHapley Additive exPlanations (SHAP) \cite{10.5555/3295222.3295230} were used to quantify the contribution of each clinical construct to the model’s predictions. A larger absolute SHAP value indicates a greater impact of the corresponding construct on the model’s output. To be specific, Tree SHAP was applied to each of the ten fine-tuned XGBoost models, and the resulting SHAP values were averaged to identify the clinical constructs that consistently played a more influential role in the model's decisions.

\subsection{A preliminary clinical evaluation of LLM explanations}
A preliminary study was conducted to assess the clinical validity of the explanations generated by LLMs. Ten samples were selected from DementiaBank, ensuring a broad distribution across the full range of MMSE scores. Each sample’s input transcript, MMSE score, and diagnosis label were provided to eight speech-language pathologists (SLPs), along with the corresponding outputs generated by Claude 3.5 Sonnet. Following their review, the SLPs were asked to rate their level of agreement with the LLM’s explanations using a 5-point Likert scale and to provide a brief rationale when they disagreed with the model output.

\section{Results}

\subsection{LLM performance on DementiaBank}

\begin{table*}[th]
\caption{Performance of the LLMs on DementiaBank using ASR-generated transcripts}
\label{tab:DB_asr}
\centering
\begin{tabular}{@{}l l l l l l l l l}
\toprule
\textbf{Construct}                        & \textbf{LLM} & \textbf{Control} & \textbf{Clinical} & \textbf{Group}      & \textbf{Hedge’s $g$} & \textbf{PCC} & \textbf{ICC} & \textbf{WSCV} \\
                                          &              & \textbf{(mean)}  & \textbf{(mean)}   & \textbf{Difference} &                    &              &              &               \\
\midrule
\textbf{Salience of information}          & Claude       & 0.87             & 1.70              & 0.95                & $-0.83^{**}$           & $-0.55^{**}$     & $0.43^{**}$      & 0.51          \\
                                          & 4o           & 1.26             & 1.90              & 0.51                & $-0.81^{**}$           & $-0.50^{**}$     & $0.32^{**}$      & 0.39          \\
                                          & 4o-mini      & 1.76             & 1.99              & 0.13                & $-0.33^{**}$           & $-0.36^{**}$     & 0.09         & 0.16          \\
                                          & LLaMA        & 1.95             & 1.92              & $-0.02$               & $-0.09$              & $-0.05$        & 0.06         & 0.50          \\
\textbf{Semantic categories}              & Claude       & 0.77             & 1.67              & 1.17                & $-1.05^{**}$           & $-0.60^{**}$     & $0.43^{**}$      & 0.45          \\
                                          & 4o           & 1.41             & 1.98              & 0.40                & $-0.77^{**}$           & $-0.50^{**}$     & $0.33^{**}$      & 0.28          \\
                                          & 4o-mini      & 1.47             & 1.86              & 0.27                & $-0.59^{**}$           & $-0.45^{**}$     & 0.15         & 0.23          \\
                                          & LLaMA        & 1.38             & 1.46              & 0.06                & $-0.10$              & $-0.09$        & 0.04         & 0.63          \\
\textbf{Referential cohesion}             & Claude       & 0.48             & 1.39              & 1.90                & $-0.82^{**}$           & $-0.53^{**}$     & $0.43^{**}$      & 0.52          \\
                                          & 4o           & 0.75             & 1.67              & 1.23                & $-0.85^{**}$           & $-0.55^{**}$     & $0.34^{**}$      & 0.61          \\
                                          & 4o-mini      & 1.51             & 1.91              & 0.26                & $-0.60^{**}$           & $-0.46^{**}$     & $0.25^{*}$       & 0.22          \\
                                          & LLaMA        & 2.04             & 2.14              & 0.05                & $-0.22$            & $-0.14^{*}$      & 0.03         & 0.49          \\
\textbf{Causal and temporal relations}    & Claude       & 1.26             & 1.93              & 0.53                & $-0.77^{**}$           & $-0.46^{**}$     & $0.36^{**}$      & 0.46          \\
                                          & 4o           & 1.93             & 2.38              & 0.23                & $-0.63^{**}$           & $-0.38^{**}$     & $0.37^{**}$      & 0.29          \\
                                          & 4o-mini      & 1.95             & 2.13              & 0.09                & $-0.47^{**}$           & $-0.33^{**}$     & 0.07         & 0.14          \\
                                          & LLaMA        & 1.67             & 1.77              & 0.06                & $-0.16$            & $-0.10$        & 0.03         & 0.58          \\
\textbf{Mental state language}            & Claude       & 1.54             & 2.12              & 0.38                & $-0.65^{**}$           & $-0.37^{**}$     & $0.40^{**}$      & 0.39          \\
                                          & 4o           & 2.25             & 2.63              & 0.17                & $-0.41^{**}$           & $-0.30^{**}$     & $0.41^{**}$      & 0.25          \\
                                          & 4o-mini      & 2.18             & 2.48              & 0.14                & $-0.38^{**}$           & $-0.24^{**}$     & $0.28^{*}$       & 0.23          \\
                                          & LLaMA        & 1.60             & 1.43              & $-0.11$               & 0.07               & $-0.01$        & 0.04         & 0.79          \\
\textbf{Structural language and speech}   & Claude       & 1.34             & 2.02              & 0.51                & $-0.84^{**}$           & $-0.51^{**}$     & $0.45^{**}$      & 0.32          \\
                                          & 4o           & 1.28             & 1.88              & 0.47                & $-0.78^{**}$           & $-0.49^{**}$     & $0.32^{**}$      & 0.37          \\
                                          & 4o-mini      & 1.46             & 1.88              & 0.29                & $-0.66^{**}$           & $-0.47^{**}$     & $0.22^{*}$       & 0.18          \\
                                          & LLaMA        & 1.76             & 1.85              & 0.05                & $-0.16$            & $-0.10$        & 0.10         & 0.68          \\
\textbf{General cognition and perception} & Claude       & 1.28             & 2.13              & 0.66                & $-0.85^{**}$           & $-0.52^{**}$     & $0.39^{**}$      & 0.45          \\
                                          & 4o           & 1.63             & 2.18              & 0.34                & $-0.60^{**}$           & $-0.42^{**}$     & $0.23^{*}$       & 0.39          \\
                                          & 4o-mini      & 1.81             & 2.10              & 0.16                & $-0.46^{**}$           & $-0.41^{**}$     & 0.01         & 0.16          \\
                                          & LLaMA        & 1.79             & 1.95              & 0.09                & $-0.26^{*}$           & $-0.15^{*}$      & 0.03         & 0.64         \\
\bottomrule
\multicolumn{9}{l}{\footnotesize $^{**}p < 0.01$, $^{*}p < 0.05$} \\
\end{tabular}
\end{table*}

We begin by comparing the discriminative ability of the severity scores generated by each LLM. As W‑ADRC contains protected health information (PHI), this analysis is conducted on DementiaBank, which does not carry PHI-related restrictions. Table \ref{tab:DB_manual} summarizes the performance of each LLM when using manually transcripts as input, whereas Table \ref{tab:DB_asr} reports performance obtained with ASR-generated transcripts. As shown in Table \ref{tab:DB_manual}, Claude 3.5 Sonnet yields the strongest discriminative performance, with medium to large effect sizes and moderate test–retest reliability. Its WSCV values are consistently lower than the corresponding percentage group differences, indicating that within-subject variability is unlikely to confound distinctions between Control and Clinical groups. Claude’s scores also show medium to large correlations with MMSE, further supporting their validity in cognitive assessment. Comparable patterns are observed on GPT-4o, which ranks second across all performance metrics. GPT-4o-mini slightly outperforms LLaMA on most constructs. While its discriminative abilities remain significant, the repeatability is limited. LLaMA-3.2-3B performs the worst overall. Compared to manual transcripts, ASR-generated transcripts generally lead to reduced performance across all LLMs, with the decline being particularly pronounced for LLaMA.

\begin{table*}[th]
\caption{Performance of the LLMs on W-ADRC using manual transcripts}
\label{tab:ADRC_manual}
\centering
\begin{tabular}{@{}l l l l l l l l l l}
\toprule
\textbf{Construct}                        & \textbf{LLM}  & \textbf{Control} & \textbf{Clinical} & \textbf{Group}      & \textbf{Hedge’s $g$} & \textbf{PCC} & \textbf{PCC} & \textbf{ICC} & \textbf{WSCV} \\
                                          &               & \textbf{(mean)}  & \textbf{(mean)}   & \textbf{Difference} &                    & \textbf{\_mmse}                   & \textbf{\_ravlt}                    &              &               \\
\midrule
\textbf{Salience of information}          & LLaMA         & 1.83             & 1.89              & 0.03                & $-0.03$              & $-0.07$              & $-0.10$               & 0.39         & 0.44          \\
                                          & LLaMA & 2.28             & 2.04              & $-0.10$               & $0.34$             & $-0.03$              & 0.03                & 0.01         & 0.41          \\
                                          & -adapted& & & & & & & &\\
                                          & LR            & 0.44             & 0.87              & 0.98                & $-0.44^{*}$           & $-0.27^{**}$           & $-0.24^{**}$            & 0.23         & 0.71          \\
\textbf{Semantic categories}              & LLaMA         & 1.48             & 1.67              & 0.12                & $-0.10$              & $-0.06$              & $-0.09$               & 0.44         & 0.56          \\
                                          & LLaMA & 2.00             & 2.02              & 0.01                & $-0.04$              & $-0.07$              & $-0.10$               & 0.06         & 0.57          \\
                                          & -adapted& & & & & & & &\\
                                          & LR            & 0.58             & 0.82              & 0.41                & $-0.34$            & $-0.22^{**}$           & $-0.17^{*}$             & 0.26         & 0.86          \\
\textbf{Referential cohesion}             & LLaMA         & 2.09             & 2.31              & 0.12                & $-0.13$              & $-0.13$              & $-0.14$               & $0.56^{*}$       & 0.37          \\
                                          & LLaMA & 2.32             & 2.38              & 0.03                & $-0.08$              & $-0.10$              & $-0.09$               & 0.24         & 0.38          \\
                                          & -adapted& & & & & & & &\\
                                          & LR            & 0.25             & 0.64              & 1.56                & $-0.55^{*}$           & $-0.25^{**}$           & $-0.20^{*}$              & 0.47         & 0.31          \\
\textbf{Causal and temporal}    & LLaMA         & 1.66             & 1.93              & 0.17                & $-0.20$              & $-0.15$              & $-0.14$               & 0.21         & 0.62          \\
\textbf{relations}                & LLaMA & 2.10             & 2.09              & $-0.01$               & 0.01               & $-0.04$              & $-0.08$               & 0.05         & 0.59          \\
                                          & -adapted& & & & & & & &\\
                                          & LR            & 0.84             & 1.24              & 0.48                & $-0.42$           & $-0.24^{**}$           & $-0.16$               & 0.11         & 0.81          \\
\textbf{Mental state language}            & LLaMA         & 1.51             & 1.69              & 0.12                & $-0.01$              & $-0.14$              & $-0.10$               & 0.12         & 0.64          \\
                                          & LLaMA & 2.67             & 2.62              & $-0.02$               & 0.19               & $-0.03$              & $-0.01$               & 0.05         & 0.17          \\
                                          & -adapted& & & & & & & &\\
                                          & LR            & 1.35             & 1.80              & 0.33                & $-0.19$              & $-0.18^{*}$             & $-0.19^{*}$              & 0.16         & 0.58          \\
\textbf{Structural language}   & LLaMA         & 1.82             & 2.02              & 0.11                & $-0.15$              & $-0.12$              & $-0.05$               & 0.28         & 0.54          \\
\textbf{and speech}                                          & LLaMA & 2.19             & 2.18              & $-0.01$               & 0.03               & $-0.01$              & $-0.04$               & 0.28         & 0.25          \\
                                          & -adapted& & & & & & & &\\
                                          & LR            & 1.15             & 1.47              & 0.28                & $-0.32$             & $-0.27^{**}$           & $-0.22^{**}$            & 0.06         & 0.26          \\
\textbf{General cognition and} & LLaMA         & 1.79             & 1.96              & 0.10                & $-0.09$              & $-0.16$              & $-0.08$               & 0.36         & 0.47          \\
\textbf{perception}                                          & LLaMA & 2.44             & 2.40              & $-0.02$               & 0.09               & $-0.03$              & $-0.03$               & 0.13         & 0.32          \\
                                          & -adapted& & & & & & & &\\
                                          & LR            & 0.82             & 1.22              & 0.49                & $-0.37$             & $-0.24^{**}$           & $-0.14$               & 0.26         & 0.62     \\
\bottomrule
\multicolumn{10}{l}{\footnotesize $^{**}p < 0.01$, $^{*}p < 0.05$, LR, logistic regression.} \\
\end{tabular}
\end{table*}

\subsection{Open-source LLM performance on W-ADRC}

We further evaluated model performance on W‑ADRC. Due to PHI restrictions, only the open-source LLaMA model was included in this analysis. As shown in Table \ref{tab:ADRC_manual}, we assessed both the out-of-the-box LLaMA and its adapted version, along with the logistic regression model trained on DementiaBank and Claude outputs. The out-of-the-box LLaMA model demonstrates limited ability to discriminate between Clinical and Control groups. Adapting LLaMA using outputs from the best-performing LLM (Claude) did not yield performance improvements and instead led to further performance degradation. In comparison, the logistic regression model exhibits substantially better performance across all constructs. However, the metric values remain relatively modest. The results obtained using ASR-generated transcripts are provided in Table \ref{tab:ADRC_asr}. Interestingly, switching to ASR-generated transcripts results in decreased performance for the logistic regression model, while enhancing the performance of both LLaMA and its adapted version, which is a phenomenon that contrasts with the observations in the previous subsection.

\begin{table*}[th]
\caption{Performance of the LLMs on W-ADRC using ASR-generated transcripts}
\label{tab:ADRC_asr}
\centering
\begin{tabular}{@{}l l l l l l l l l l}
\toprule
\textbf{Construct}               & \textbf{LLM} & \textbf{Control} & \textbf{Clinical} & \textbf{Group}      & \textbf{Hedge’s $g$} & \textbf{PCC} & \textbf{PCC} & \textbf{ICC} & \textbf{WSCV} \\
                                 &              & \textbf{(mean)}  & \textbf{(mean)}   & \textbf{Difference} &                    & \textbf{\_mmse}                   & \textbf{\_ravlt}                    &              &               \\
\midrule
\textbf{Salience of information} & LLaMA        & 1.71             & 2.00              & 0.16                & $-0.30$            & -0.16              & $-0.18^{*}$             & 0.18         & 0.43          \\
                                 & LLaMA        & 1.68             & 1.76              & 0.04                & $-0.05$              & $-0.05$              & $-0.13$               & 0.18         & 0.51          \\
                                 & -adapted     &                  &                   &                     &                    &                    &                     &              &               \\
                                 & LR           & 0.14             & 0.47              & 2.31                & $-0.50^{*}$           & $-0.18^{*}$            & $-0.23^{**}$            & 0.13         & 0.47          \\
\textbf{Semantic categories}     & LLaMA        & 1.22             & 1.58              & 0.30                & $-0.30$            & $-0.15$              & $-0.10$               & 0.03         & 0.59          \\
                                 & LLaMA        & 1.61             & 1.96              & 0.20                & $-0.26$            & $-0.13$              & $-0.18^{*}$             & 0.36         & 0.23          \\
                                 & -adapted     &                  &                   &                     &                    &                    &                     &              &               \\
                                 & LR           & 0.45             & 0.64              & 0.42                & $-0.33$            & $-0.20^{*}$            & $-0.18^{*}$             & 0.13         & 0.63          \\
\textbf{Referential cohesion}    & LLaMA        & 1.98             & 2.38              & 0.20                & $-0.26$            & $-0.18^{*}$            & $-0.18^{*}$             & 0.23         & 0.53          \\
                                 & LLaMA        & 1.84             & 2.09              & 0.14                & $-0.24$            & $-0.09$              & $-0.21^{**}$            & 0.17         & 0.30          \\
                                 & -adapted     &                  &                   &                     &                    &                    &                     &              &               \\
                                 & LR           & 0.13             & 0.15              & 0.15                & $-0.16$              & $-0.02$              & $-0.03$               & 0.07         & 0.31          \\
\textbf{Causal and temporal}     & LLaMA        & 1.48             & 1.80              & 0.21                & $-0.25$            & $-0.14$              & $-0.13$               & 0.09         & 0.58          \\
\textbf{relations}               & LLaMA        & 1.70             & 1.98              & 0.17                & $-0.14$              & $-0.13$              & $-0.23^{**}$            & 0.01         & 0.46          \\
                                 & -adapted     &                  &                   &                     &                    &                    &                     &              &               \\
                                 & LR           & 0.56             & 0.98              & 0.75                & $-0.38$            & $-0.20^{*}$            & $-0.22^{**}$            & 0.24         & 0.97          \\
\textbf{Mental state language}   & LLaMA        & 1.44             & 1.73              & 0.20                & $-0.30$            & $-0.07$              & $-0.08$               & 0.15         & 0.58          \\
                                 & LLaMA        & 2.11             & 2.31              & 0.09                & $-0.11$              & $-0.06$              & $-0.19^{*}$             & 0.31         & 0.33          \\
                                 & -adapted     &                  &                   &                     &                    &                    &                     &              &               \\
                                 & LR           & 1.19             & 1.49              & 0.25                & $-0.09$              & $-0.16$              & $-0.18^{*}$             & 0.33         & 0.42          \\
\textbf{Structural language}     & LLaMA        & 1.60             & 2.13              & 0.32                & $-0.42^{*}$           & $-0.20^{*}$            & $-0.14$               & 0.12         & 0.59          \\
\textbf{and speech}              & LLaMA        & 1.92             & 2.24              & 0.17                & $-0.22$              & $-0.18^{*}$            & $-0.20^{*}$             & 0.10         & 0.31          \\
                                 & -adapted     &                  &                   &                     &                    &                    &                     &              &               \\
                                 & LR           & 0.98             & 1.18              & 0.21                & $-0.20$              & $-0.26^{**}$           & $-0.16$               & 0.23         & 0.10          \\
\textbf{General cognition and}   & LLaMA        & 1.61             & 2.04              & 0.26                & $-0.30$            & $-0.19^{*}$            & $-0.14$               & 0.19         & 0.65          \\
\textbf{perception}              & LLaMA        & 2.03             & 2.31              & 0.14                & $-0.18$              & $-0.16$              & $-0.22^{**}$            & 0.32         & 0.24          \\
                                 & -adapted     &                  &                   &                     &                    &                    &                     &              &               \\
                                 & LR           & 0.28             & 0.80              & 1.86                & $-0.70^{**}$           & $-0.32^{**}$           & $-0.28^{**}$            & 0.07         & 0.47         \\
\bottomrule
\multicolumn{10}{l}{\footnotesize $^{**}p < 0.01$, $^{*}p < 0.05$, LR, logistic regression.} \\
\end{tabular}
\end{table*}

\subsection{Performance of machine learning-based dementia detection}

\begin{figure*}[t]
  \centering
  \includegraphics[width=\linewidth]{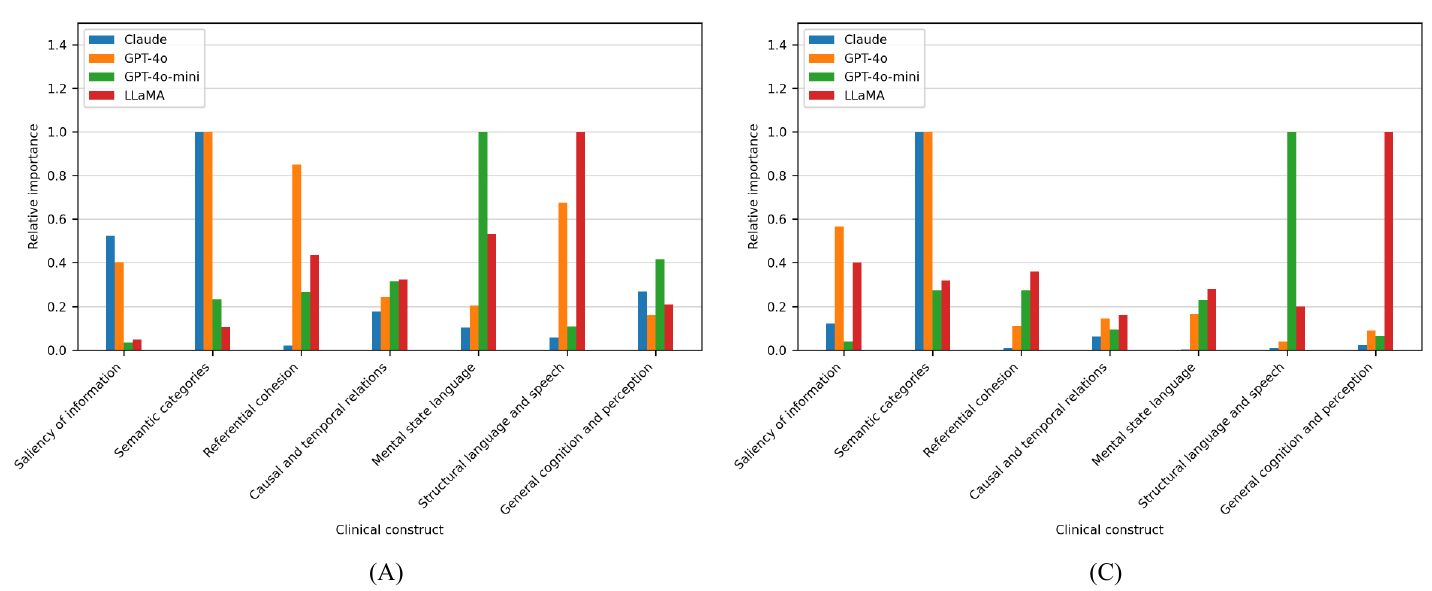}
  \caption{Normalized relative importance of each construct in the XGBoost model’s decision-making process. The model was trained using severity scores derived from (A) manual transcripts and (B) ASR-generated transcripts.}
  \label{fig:shap}
\end{figure*}

As shown in Table \ref{tab:xgb_manual}, severity scores produced by Claude 3.5 Sonnet and GPT-4o effectively support XGBoost in distinguishing between dementia patients and healthy controls. Claude achieves the highest performance, with an accuracy of 85\% and a corresponding F1 score of 0.84. In contrast, the severity scores generated by GPT-4o-mini and LLaMA yield suboptimal performance, with accuracies around 0.6. The results obtained using ASR-generated transcripts are provided in Table \ref{tab:xgb_asr}. Severity scores derived from ASR-generated transcripts generally yielded lower dementia detection accuracy. This finding align with the trends observed in Tables \ref{tab:DB_asr}. We further examined the relative importance of the clinical constructs in the models’ decision-making processes. As shown in Figure \ref{fig:shap}, the highest-performing models frequently relied on “Saliency of information” and “Semantic categories” for their predictions, while “Causal and temporal relations” and “Mental state language” contributed the least in almost all cases.

\begin{table}[th]
\caption{Dementia classification performance of the LLMs on ADReSS dataset using manual transcripts}
\label{tab:xgb_manual}
\centering
\begin{tabular}{l l l l l}
\toprule
\textbf{LLM}     & \textbf{Accuracy} & \textbf{Precision} & \textbf{Recall} & \textbf{F1}       \\
\midrule
\textbf{Claude}  & 0.85              & 0.90               & 0.79            & 0.84              \\
\textbf{4o}      & 0.81              & 0.83               & 0.79            & 0.81              \\
\textbf{4o-mini} & 0.58              & 0.61               & 0.46            & 0.52              \\
\textbf{LLaMA}   & 0.63              & 0.64               & 0.58            & 0.61              \\
\bottomrule
\end{tabular}
\end{table}

\begin{table}[th]
\caption{Dementia classification performance of the LLMs on ADReSS dataset using ASR-generated transcripts}
\label{tab:xgb_asr}
\centering
\begin{tabular}{l l l l l}
\toprule
\textbf{LLM}     & \textbf{Accuracy} & \textbf{Precision} & \textbf{Recall} & \textbf{F1}       \\
\midrule
\textbf{Claude}  & 0.84              & 0.89               & 0.77            & 0.83              \\
\textbf{4o}      & 0.78              & 0.75               & 0.82            & 0.78              \\
\textbf{4o-mini} & 0.64              & 0.62               & 0.73            & 0.67              \\
\textbf{LLaMA}   & 0.51              & 0.50               & 0.45            & 0.48              \\
\bottomrule
\end{tabular}
\end{table}

\subsection{Evaluation of LLM explanations}
Table \ref{tab:agreement} summarizes the distribution of SLP ratings for each sample. Across all SLPs, the mean agreement score was 3.99 ± 0.92 out of 5, indicating a medium-to-high level of alignment between the LLM-generated explanations and clinicians’ evaluations. In addition, an ICC of 0.63 was obtained based on ratings provided by eight SLPs, reflecting a moderate level of inter-rater variability among the SLPs.

\begin{table}[th]
\caption{SLP agreement ratings for Claude evaluations}
\label{tab:agreement}
\centering
\begin{tabular}{l l l l}
\toprule
\textbf{Sample number} & \textbf{Diagnosis} & \textbf{MMSE} & \textbf{Average} \\
 &  &  & \textbf{agreement} \\
\midrule
\textbf{1}               & Probable AD        & 3             & 3.63 ±   1.06                  \\
\textbf{2}               & Probable AD        & 8             & 4.75 ± 0.46                    \\
\textbf{3}               & Probable AD        & 15            & 3.63 ±   1.19                  \\
\textbf{4}               & Probable AD        & 16            & 3.63 ± 0.74                    \\
\textbf{5}               & Probable AD        & 22            & 3.25 ±   1.28                  \\
\textbf{6}               & Probable AD        & 23            & 3.88 ± 0.64                    \\
\textbf{7}               & Control            & 30            & 4.25 ±   0.71                  \\
\textbf{8}               & Control            & 30            & 4.38 ± 0.74                    \\
\textbf{9}               & Control            & 29            & 4.13 ±   0.64                  \\
\textbf{10}              & Control            & 27            & 4.38 ± 0.74                    \\
\bottomrule
\end{tabular}
\end{table}

\section{Discussion}

This study introduced seven clinical constructs tailored to the Cookie Theft picture description task and prompted several LLMs to assess each construct by assigning severity scores and generating example-based explanations. As presented in Table \ref{tab:DB_manual}, Claude 3.5 Sonnet outperforms other LLMs in all metrics, indicating that advanced LLMs hold promise for effectively evaluating the cognitive-linguistic abilities across a range of cognitive statuses. This finding is further solidified by the results on ADReSS (Table \ref{tab:xgb_manual}), where severity scores generated by Claude achieved an accuracy of 85\%, exceeding the 81.3\% accuracy reported in prior LLM-based studies \cite{botelho24_interspeech, zheng2024alzheimer}. Although this performance remains lower than that of traditional BERT-based studies \cite{wang22l_interspeech}, it is important to note that LLMs can directly operationalize clinically meaningful constructs into quantitative measures and provide explanations for their decisions, offering a significant advantage for practical applications compared to black-box models like BERT. A preliminary evaluation of Claude’s outputs yields an average agreement score of 3.99 out of 5, indicating a medium-to-high level of agreement between the LLM and the SLPs. However, the moderate inter-rater variability highlights differences in expert opinion regarding the clinical utility of the scores and explanations. SLPs’ comments reveal several issues: (1) The definitions of certain constructs (“Saliency of information”, “General cognition and perception”) may lack sufficient clarity for consistent interpretation by the LLM. Sometimes the LLM failed to adhere to the intended scope and evaluated aspects falling under others, resulting in overly severe assessments. (2) The LLM exhibited a relatively low tolerance for informal language usage (e.g. “outta”, “hafta” or elliptical expressions), and considered them as signs of impairment, leading to inflated scores in constructs like “Semantic categories”, “Referential cohesion”, “Structural language and speech”, and “General cognition and perception”. (3) The LLM demonstrated difficulty in accurately identifying symptoms associated with “Causal and temporal relations”, occasionally resulting in overly lenient evaluations. Prior evaluations have shown that certain versions of ChatGPT are prone to hallucinations when identifying causal relationships \cite{gao2023chatgpt}, and similar limitations may also exist in Claude. (4) For “Mental state language”, the LLM evaluations were considered too severe. Several SLPs noted that the standard task prompt (e.g., “tell me everything you see in this picture”) does not explicitly request participants to interpret characters’ mental states. Therefore, the absence of mental state language in a description may not necessarily be interpreted as an impairment. These observations offer valuable insights for refining the evaluation framework and the prompt design in future iterations.

LLaMA shows the weakest performance on both DementiaBank (Table \ref{tab:DB_manual}) and W-ADRC (Table \ref{tab:ADRC_manual}), likely due to its smaller size compared to proprietary LLMs. Despite being lightweight, it still demands considerable resources for local use. These results underscore the challenges of deploying LLMs in PHI-restricted clinical settings. Specific to LLaMA’s low performance on W-ADRC, another contributing factor could be sample distribution. The Clinical group in W-ADRC shows significantly milder symptoms compared to those in DementiaBank, as reflected by a much higher average MMSE score in Table \ref{tab:dataset}. Consequently, the group differences in W-ADRC are less pronounced, which may partially account for LLaMA’s reduced performance on the W-ADRC task. Regarding the adaptation strategies, as shown in Table \ref{tab:ADRC_manual}, adapting LLaMA using DementiaBank did not improve its performance, suggesting that simply adding data without optimizing the training approach may be insufficient. In contrast, training a logistic regression model to estimate each severity score resulted in significantly better performance, highlighting the importance of task-specific training. Considering the model was trained using Claude-generated severity scores as ground-truth labels, this finding further validates the utility of the severity scores produced by Claude.

At the construct level, both the statistical analysis and SHAP study indicate that “Mental state language” and “Causal and temporal relations” are less effective than other constructs. For “Mental state language”, we notice that even the best-performing model (Claude) frequently assigned a score of “2” to healthy controls, thereby reducing the group difference. This finding is consistent with SLP feedback, which noted that some participants omitted mental state language simply because they were not prompted to include it. The LLMs, however, misinterpreted this omission as indicative of impairment. For “Causal and temporal relations”, the LLMs tended to assign scores of “1” or “2” to clinical speakers. This observation again aligns with SLP feedback, which noted that Claude sometimes failed to detect severe impairments in a speaker’s ability to maintain causal relations and sequence of events. Previous studies have shown that LLMs like ChatGPT are inclined to assume causal relations between events \cite{gao2023chatgpt}. This tendency may result in underestimating impairment severity and generating scores with reduced discriminative power.

Switching from manual to ASR-generated transcripts reduced performance across all LLMs on DementiaBank but improved LLaMA’s performance on W-ADRC. This discrepancy likely stems from differences in transcription quality. DementiaBank’s older, lower-quality recordings impaired ASR accuracy, obscuring key cognitive-linguistic symptoms, while W-ADRC’s clearer recordings allowed the ASR system to accurately preserve clinical information, leading to a better performance. This quality gap also explains why logistic regression performance dropped when switching to ASR. As described in Methods, the model was trained on DementiaBank and tested on W-ADRC. Manual transcripts ensured comparable quality across training and test sets, supporting effective severity scoring. In contrast, ASR-generated transcripts introduced a quality mismatch, challenging the model’s generalizability and leading to reduced performance. 

\section{Conclusion and future work}

In this study, we prompted several LLMs to operationalize seven task-specific clinical constructs and generate example-based explanations. Results highlight the potential of advanced models like Claude 3.5 Sonnet to assess cognitive status from Cookie Theft picture descriptions, with preliminary SLP feedback indicating moderately high agreement with model-generated scores and rationales. However, limitations remain. First, only one lightweight LLaMA model was evaluated, which is insufficient to fully assess the capability of local LLMs. Future work should extend this analysis to more advanced and larger-scale local models. Second, the validity of the generated explanations was only assessed through a simple preliminary experiment. Given the importance of interpretability in clinical context, another comprehensive evaluation is necessary to rigorously assess the accuracy and utility of these explanations. Finally, improvements to the framework are needed, such as fine-tuning the ASR tool to enhance accuracy while preserving clinically relevant symptoms, refining prompts with well-defined constructs and instructions, and exploring more effective few-shot example selection strategies.

\section{Acknowledgments}
This work was supported in part by NIH-NIA Grant 5R01AG082052, 1T32AG082658-01A1, and funding from the John and Tami Marick Family Foundation.
We gratefully acknowledge Joshua Breger, Phoebe Crumpton, Elena Groves, Madeline Hale, Kristin Murphey, Iris Nowenstein, Ileana Ratiu, and Elizabeth Trueba for their support in evaluating LLM outputs. We also acknowledge Sanjay Asthana, Sterling Johnson, and the participants and study teams from the Pittsburgh and Wisconsin Alzheimer’s Disease Research Centers.

\section{Generative AI Use Disclosure}
Generative AI tools were used solely for language editing and polishing of the manuscript.

\bibliographystyle{IEEEtran}
\bibliography{mybib}

\end{document}